# What quantum "state" really is?

*Alexander M. SOIGUINE*

**Abstract:** To get out of logical deadlock in interpreting gedanken experiments like "Schrodinger cat", actual meaning of a "wave function", or a "state", in the case of complex two-dimensional Hilbert space, is shown to be an element of even subalgebra of geometric algebra [1], [2] over three-dimensional Euclidian space.

ב. לֹא תִהְיֶה אַחֲרֵי רַבִּים לְרָעֹת וְלֹא תַעֲנֶה עַל
רִב לִנְטֹת אַחֲרֵי רַבִּים לְהַטֹּת:

Exodus 23:2[1]

1. Introduction

The geometric algebra $G_3^+$ [1] formalism is more profound compared to formal quantum mechanical representation of two-state observables by "complex" amplitudes as components of elements in two-dimensional Hilbert space. It has been used in to describe tossed coin experiment that was pretty similar to traditional quantum mechanical formalism [3].

Mathematical frame will be here the even subalgebra $G_3^+$ of elements:

$$so(\alpha, \beta, S) = \alpha + \beta I_S$$

of geometric algebra $G_3$ over Euclidian space $E_3$. $G_3$ has the basis

$$\{1, e_1, e_2, e_3, e_1e_2, e_2e_3, e_3e_1, e_1e_2e_3\},$$

---

[1] You shall not follow the majority for evil, and you shall not respond concerning a lawsuit to follow many to pervert [justice].

where $\{e_i\}$ are orthonormal basis vectors in $E_3$, $\{e_i e_j\}$ - oriented, mutually orthogonal unit value areas spanned by $e_i$ and $e_j$ as edges, $e_1 e_2 e_3$ - unit value oriented volume spanned by ordered edges $e_1$, $e_2$ and $e_3$.

Subalgebra $G_3^+$ is spanned by $\{1, e_1 e_2, e_2 e_3, e_3 e_1\}$. Variables $\alpha$ and $\beta$ in $so(\alpha, \beta, S)$ are (real[2]) scalars, $I_S$ is a unit size oriented area (lefthanded or righthanded) in an arbitrary given plane $S \subset E_3$.

I explained in detail [4], [5] that elements $so(\alpha, \beta, S) = \alpha + \beta I_S$ only differ from what is traditionally called "complex numbers" by the fact that $S \subset E_3$ is an arbitrary, variable plane and is not the whole space of game. Putting it simply, $\alpha + \beta I_S$ are "complex numbers" depending on $E_2$ embedded into $E_3$. $E_2$ is the space where $S$ belongs. Traditional "imaginary unit" $i$ is just $I_S$ when it is not necessary to specify the plane – everything is going on in one fixed plane, not in 3D world. Fully formal way of using $i$ as a "number", just with additional algebraic property $i^2 = -1$, may be a source of deeply wrong interpretations, particularly in quantum mechanics.

## 2. Rotations and elements of $G_3^+$

If $g_3$ is element of $G_3$, its rotation in 3D can be written as [1], [2]:

$$g_3 \rightarrow (\alpha - \beta I_S) g_3 (\alpha + \beta I_S),$$

where $\alpha + \beta I_S = \alpha + \beta(b_1 e_2 e_3 + b_2 e_3 e_1 + b_3 e_1 e_2)$ is unit value element of $G_3^+$, that's $\alpha^2 + \beta^2 = 1$ and $b_1^2 + b_2^2 + b_3^2 = 1$. I will name such elements $G_3^+$ - states. $G_3^+$ - states can also be considered as points on unit radius sphere $S^3$. Parameter $\alpha$ is cosine of half of angle of rotation, $I_S$ is unit value bivector in plane $S$ orthogonal to axis of rotation[3]. Since $\alpha^2 + \beta^2 = 1$ we can write

$$g_3 \rightarrow (\cos\frac{\varphi}{2} - \sin\frac{\varphi}{2} I_S) g_3 (\cos\frac{\varphi}{2} + \sin\frac{\varphi}{2} I_S) \equiv \exp\left(-\frac{\varphi}{2} I_S\right) g_3 \exp\left(\frac{\varphi}{2} I_S\right)$$

Particularly, if $g_3$ is a bivector, $C = C_1 e_2 e_3 + C_2 e_3 e_1 + C_3 e_1 e_2$ [4] (representing, for example, object with axial symmetry), its rotation has symmetry:

---

[2] Scalars should always be real. "Complex" scalars are not scalars.
[3] Axis of rotation together with the bivector orientation should be compatible with orientation of $e_1 e_2 e_3$.
[4] I will often identically use the same term for bivector and its plane.

$$C \to \exp\left(-\frac{\varphi}{2}I_S\right)C\exp\left(\frac{\varphi}{2}I_S\right) = \exp\left(-\frac{\varphi}{2}I_S\right)\exp\left(-\frac{\vartheta}{2}I_C\right)C\exp\left(\frac{\vartheta}{2}I_C\right)\exp\left(\frac{\varphi}{2}I_S\right),$$

that simply means that nothing changes if bivector is also rotated in its own plane [3]. It formally follows from:

$$(\alpha - \beta I_S)C(\alpha + \beta I_S) = (\alpha^2 - \beta^2)C - 2\alpha\beta(I_S \times C) - 2\beta^2 I_S(I_S \cdot C), \qquad (2.1)$$

where $I_S \times C = \frac{1}{2}(I_S C - C I_S)$, $I_S \cdot C = \frac{1}{2}(I_S C + C I_S)$. That obviously means that if we rotate bivector with unit element $so(\alpha, \beta, S)$ and both have parallel bivector planes, that's $I_S = C$, then bivector $C$ remains the same.

If we observe a real physical object identified by bivector, say axially symmetric object, and the result of observation is just its orientation in 3D, its rotation in its own plane around central axis is not detectable. Angle, or phase, of such rotation should be considered as unspecified variable [3]. But that does not mean the variable does not comprise some physically meaningful feature.

If $C = e_2 e_3$ then:

$$(\alpha - \beta I_S)e_2 e_3(\alpha + \beta I_S) = (\alpha - \beta(b_1 e_2 e_3 + b_2 e_3 e_1 + b_3 e_1 e_2))e_2 e_3 (\alpha + \beta(b_1 e_2 e_3 + b_2 e_3 e_1 + b_3 e_1 e_2)) =$$

$$(\alpha - \beta_1 e_2 e_3 - \beta_2 e_3 e_1 - \beta_3 e_1 e_2)e_2 e_3(\alpha + \beta_1 e_2 e_3 + \beta_2 e_3 e_1 + \beta_3 e_1 e_2) =$$

$$(\alpha^2 + \beta_1^2 - \beta_2^2 - \beta_3^2)e_2 e_3 + 2(\alpha\beta_3 + \beta_1\beta_2)e_3 e_1 + 2(\beta_1\beta_3 - \alpha\beta_2)e_1 e_2, \text{ where } \beta_i \equiv \beta b_i.$$

This is classical Hopf fibration $S^3 \to S^2 : so(\alpha, \beta, S) \xrightarrow{e_2 e_3} C_{rot}$, where

$$C_{rot} = (\alpha^2 + \beta_1^2 - \beta_2^2 - \beta_3^2)e_2 e_3 + 2(\alpha\beta_3 + \beta_1\beta_2)e_3 e_1 + 2(\beta_1\beta_3 - \alpha\beta_2)e_1 e_2 \qquad (2.2)$$

Also it shows that "imaginary" number $i$ in that Hopf fibration should geometrically be unit value oriented area $e_2 e_3$ orthogonal to basis vector $e_1$. Orientations of $e_1$ and $e_2 e_3$ must be in agreement with $e_1 e_2 e_3$ orientation.

Generalized Hopf fibration can be generated by any bivector $C$:

$$S^3 \to S^2 : so(\alpha, \beta, S) \xrightarrow{C} C_{rot},$$

and (2.1) gives the result.

Often it is more convenient to deal with this result explicitly written in the basis bivector components. We know the result (2.2) for $so(\alpha,\beta,S) \xrightarrow{e_2e_3} C_{rot}$. For two other basis bivector components we have:

$$so(\alpha,\beta,S) \xrightarrow{e_3e_1} C_{rot} = 2(\beta_1\beta_2 - \alpha\beta_3)e_2e_3 + [(\alpha^2 + \beta_2^2) - (\beta_1^2 + \beta_3^2)]e_3e_1 + 2(\alpha\beta_1 + \beta_2\beta_3)e_1e_2$$

$$so(\alpha,\beta,S) \xrightarrow{e_1e_2} C_{rot} = 2(\alpha\beta_2 + \beta_1\beta_3)e_2e_3 + 2(\beta_2\beta_3 - \alpha\beta_1)e_3e_1 + [(\alpha^2 + \beta_3^2) - (\beta_1^2 + \beta_2^2)]e_1e_2$$

Then:

$$so(\alpha,\beta,S) \xrightarrow{C_1e_2e_3 + C_2e_3e_1 + C_3e_1e_2} C_{rot} =$$

$$C_1\big([(\alpha^2 + \beta_1^2) - (\beta_2^2 + \beta_3^2)]e_2e_3 + 2(\alpha\beta_3 + \beta_1\beta_2)e_3e_1 + 2(\beta_1\beta_3 - \alpha\beta_2)e_1e_2\big) +$$

$$C_2\big(2(\beta_1\beta_2 - \alpha\beta_3)e_2e_3 + [(\alpha^2 + \beta_2^2) - (\beta_1^2 + \beta_3^2)]e_3e_1 + 2(\alpha\beta_1 + \beta_2\beta_3)e_1e_2\big) +$$

$$C_3\big(2(\alpha\beta_2 + \beta_1\beta_3)e_2e_3 + 2(\beta_2\beta_3 - \alpha\beta_1)e_3e_1 + [(\alpha^2 + \beta_3^2) - (\beta_1^2 + \beta_2^2)]e_1e_2\big) =$$

$$\big(C_1[(\alpha^2 + \beta_1^2) - (\beta_2^2 + \beta_3^2)] + 2C_2(\beta_1\beta_2 - \alpha\beta_3) + 2C_3(\alpha\beta_2 + \beta_1\beta_3)\big)e_2e_3 +$$

$$\big(2C_1(\alpha\beta_3 + \beta_1\beta_2) + C_2[(\alpha^2 + \beta_2^2) - (\beta_1^2 + \beta_3^2)] + 2C_3(\beta_2\beta_3 - \alpha\beta_1)\big)e_3e_1 +$$

$$\big(2C_1(\beta_1\beta_3 - \alpha\beta_2) + 2C_2(\alpha\beta_1 + \beta_2\beta_3) + C_3[(\alpha^2 + \beta_3^2) - (\beta_1^2 + \beta_2^2)]\big)e_1e_2 \qquad (2.3)$$

*Remark 2.1*

If we take bivector expanded not in $\{e_2e_3, e_3e_1, e_1e_2\}$ but in arbitrary basis $\{B_1, B_2, B_3\}$, satisfying the same multiplication rules $B_1B_2 = -B_3, B_1B_3 = B_2, B_2B_3 = -B_1$, that's $\alpha + \beta I_S = \alpha + \beta(b_1B_1 + b_2B_2 + b_3B_3) = \alpha + \beta_1B_1 + \beta_1B_1 + \beta_1B_1, \beta_i = \beta b_i$, then, for example, the result of rotation of $B_1$ is:

$$(\alpha^2 + \beta_1^2 - \beta_2^2 - \beta_3^2)B_1 + 2[(\alpha\beta_3 + \beta_1\beta_2)B_2 - (\alpha\beta_2 - \beta_1\beta_3)B_3],$$

exactly the same as in the $\{e_2e_3, e_3e_1, e_1e_2\}$ case. So, (2.3) remains valid when replacing $\{e_2e_3, e_3e_1, e_1e_2\}$ with $\{B_1, B_2, B_3\}$.

## 3. $G_3^+$ - states instead of wave functions

### 3.1. Two component quantum states and $G_3^+$ - states

Let's firstly establish clear relations between the two state wave functions $|\psi\rangle = (c_1, c_2)^T$, where the components $c_1, c_2$ of the column are "complex" numbers, and elements $so(\alpha, \beta, S) \equiv \alpha + \beta I_S = \alpha + \beta(b_1 e_2 e_3 + b_2 e_3 e_1 + b_3 e_1 e_2)$, $\alpha^2 + \beta^2 = 1$, $b_1^2 + b_2^2 + b_3^2 = 1$.

$G_3^+$ - states $so(\alpha, \beta, S) \equiv \alpha + \beta I_S$ are elements of $S^3$. Common approach of treating $S^3$, which sits in $R^4$, is identification of $R^4$ with $C^2$, two dimensional space of "complex" numbers, that is based on their equal dimensionalities:

$$R^4 \supset S^3 = \{z = (z_1, z_2) \in C^2; |z|^2 = \bar{z}_1 z_1 + \bar{z}_2 z_2 = 1\} \quad (3.1)$$

We are considering elements of $S^3$ as unit value $G_3^+$ - states $\alpha + \beta I_S, \alpha^2 + \beta^2 = 1$, and should strictly follow the requirement that, if "complex" numbers are involved somewhere, their plane must be explicitly defined. When (3.1) is formally used, a tacit common quantum mechanical assumption is that $z_1$ and $z_2$ have the same "complex" plane, otherwise complex conjugation is not well defined [5]. Let's make all that unambiguously clear.

Take a $G_3^+$ - state:

$$so(\alpha, \beta, S) \equiv so(\alpha, \vec{\beta}) = \alpha + \beta_1 e_2 e_3 + \beta_2 e_3 e_1 + \beta_3 e_1 e_2, \ \beta_i \equiv \beta b_i$$

We want to rewrite it as a couple of "complex" numbers, used in (3.1), with an explicitly defined plane. I will initially take a "complex" plane as one of the spanned by $\{e_2, e_3\}$, $\{e_3, e_1\}$ or $\{e_1, e_2\}$.

For $\{e_2, e_3\}$ we get:

$$so(\alpha, \vec{\beta}) = (\alpha + \beta_1 e_2 e_3) + \beta_2 e_2 e_3 e_1 e_2 + \beta_3 e_1 e_2 = (\alpha + \beta_1 e_2 e_3) + (\beta_3 + \beta_2 e_2 e_3) e_1 e_2 \equiv z_1^{2,3} + z_2^{2,3} e_1 e_2$$

For $\{e_3, e_1\}$:

$$so(\alpha, \vec{\beta}) = (\alpha + \beta_2 e_3 e_1) + \beta_3 e_3 e_1 e_2 e_3 + \beta_1 e_2 e_3 = (\alpha + \beta_2 e_3 e_1) + (\beta_1 + \beta_3 e_3 e_1) e_2 e_3 \equiv z_1^{3,1} + z_2^{3,1} e_2 e_3$$

And for $\{e_1, e_2\}$:

$$so(\alpha, \vec{\beta}) = (\alpha + \beta_3 e_1 e_2) + \beta_1 e_1 e_2 e_3 e_1 + \beta_2 e_3 e_1 = (\alpha + \beta_3 e_1 e_2) + (\beta_2 + \beta_1 e_1 e_2) e_3 e_1 \equiv z_1^{1,2} + z_2^{1,2} e_3 e_1$$

*Remark 3.1.1*

One can notice that the second members in all three cases can be written in two ways. For example:

$$z_2^{2,3} e_1 e_2 = (\beta_3 + \beta_2 e_2 e_3) e_1 e_2 = (\beta_2 - \beta_3 e_2 e_3) e_3 e_1.$$

It can be checked that geometrically both give the same bivector.

So, we have three basic maps $so(\alpha, \beta, S) \xleftrightarrow{e_i e_j} (z_1^{i,j}, z_2^{i,i})$ defined by "complex" planes $\{e_i, e_j\}$. If we are speaking about identification of $S^3$ and $C^2$, we need to explicitly say which plane we use, particularly, which of the three basis (in 3D) "complex" planes we are working in.

The "complex" plane can be different from any of basis planes $\{e_i e_j\}$. Instead of basis of three bivectors $\{e_2 e_3, e_3 e_1, e_1 e_2\}$ we can take, as above, $\{B_1, B_2, B_3\}$. If $S^3 \ni so(\alpha, \beta, S) \equiv \alpha + \beta I_S$ is expanded in basis $\{B_1, B_2, B_3\}$:

$$\alpha + \beta I_S = \alpha + \beta(b_1 B_1 + b_2 B_2 + b_3 B_3) = \alpha + \beta_1 B_1 + \beta_1 B_1 + \beta_1 B_1, \beta_i = \beta b_i,$$

then, for example, taking $B_1$ as "complex" plane we get:

$$\alpha + \beta_1 B_1 + \beta_1 B_1 + \beta_1 B_1 = \alpha + \beta_1 B_1 + \beta_2 B_1 B_3 + \beta_3 B_3 = \alpha + \beta_1 B_1 + (\beta_3 + \beta_2 B_1) B_3 \to$$
$$((\alpha + \beta_1 B_1), (\beta_3 + \beta_2 B_1)) \equiv (z_1, z_2)$$

Let's get back to the case of "complex" plane spanned by $\{e_2, e_3\}$:

$$so(\alpha, \vec{\beta}) = (\alpha + \beta_1 e_2 e_3) + \beta_2 e_2 e_3 e_1 e_2 + \beta_3 e_1 e_2 = (\alpha + \beta_1 e_2 e_3) + (\beta_3 + \beta_2 e_2 e_3) e_1 e_2 \equiv z_1^{2,3} + z_2^{2,3} e_1 e_2,$$

In this case wave function $|\psi\rangle = \left(z_1^{2,3}, z_2^{2,3}\right)^T$ corresponding to $so(\alpha, \beta, S)$ is:

$$|\psi\rangle = (\alpha + \beta_1 e_2 e_3, \beta_3 + \beta_2 e_2 e_3)^T \qquad (3.2)$$

Let's consider the Pauli's matrices[5], basis of "observables" in traditional quantum mechanics:

$$\hat{\sigma}_1 = \begin{pmatrix} 1 & 0 \\ 0 & -1 \end{pmatrix}, \hat{\sigma}_2 = \begin{pmatrix} 0 & 1 \\ 1 & 0 \end{pmatrix}, \hat{\sigma}_3 = \begin{pmatrix} 0 & i \\ -i & 0 \end{pmatrix}, \quad (i \equiv e_2 e_3)$$

---

[5] Some modifications were necessary to follow handedness.

which are in one-to-one correspondence with our basis bivectors, namely all $G_3^+$ geometric algebra equations can be expressed in terms of $\hat{\sigma}_k$ through $e_2 e_3 = I_3 \hat{\sigma}_1$, $e_3 e_1 = I_3 \hat{\sigma}_2$ and $e_1 e_2 = I_3 \hat{\sigma}_3$, where $I_3 \equiv e_1 e_2 e_3$ [4].

Usual quantum mechanical calculations with Pauli's matrices give the following:

$$\langle \psi | \hat{\sigma}_1 | \psi \rangle = \alpha^2 + \beta_1^2 - \beta_2^2 - \beta_3^2, \; \langle \psi | \hat{\sigma}_2 | \psi \rangle = 2(\alpha \beta_3 + \beta_1 \beta_2), \; \langle \psi | \hat{\sigma}_3 | \psi \rangle = 2(\beta_1 \beta_3 - \alpha \beta_2),$$

where $\langle \psi |$ is element conjugated to $|\psi\rangle$:

$$\langle \psi | = (\alpha - \beta_1 e_2 e_3, \beta_3 - \beta_2 e_2 e_3)$$

Each of the products $\langle \psi | \hat{\sigma}_i | \psi \rangle$ gives only one component of usual Hopf fibration, contrary to $(\alpha - \beta I_S) e_2 e_3 (\alpha + \beta I_S)$, giving full vector. It looks like in the mapping $so(\alpha, \beta, S) \in G_3^+ \sim S^3 \to C^2 \ni |\psi\rangle$ some information was lost. Indeed, information was lost in several ways. Only one particular basis plane remains in $|\psi\rangle$ as "complex" plane. Then, algebraic coincidence of multiplication rules for Pauli matrices and basis bivectors $\{e_i e_j\}$ is not equivalent to their geometrical identity: the matrices do not have an appropriate geometrical sense. And also the products $\langle \psi | \hat{\sigma}_i | \psi \rangle$ are scalar products of two vectors, where $|\psi\rangle$ contains complex conjugation relative to plane which is preselected as "complex" one. The whole design of $\langle \psi | \hat{\sigma}_i | \psi \rangle$ was aimed at receiving three components of rotation $(\alpha - \beta I_S) e_2 e_3 (\alpha + \beta I_S)$, without actual geometrical sense of operations. Not surprising that it was possible to formally prove [6] that "hidden variables" do not exist in the Hilbert space formalism.

Following the paradigm of [3] I assume that actual meaning of quantum mechanical wave functions, at least in the case of axially symmetric objects, is that the functions should be elements of $G_3^+$, $(\alpha + \beta I_S)$, returning observation of bivectors $C$ representing orientation of physical objects:

$$C \to (\alpha - \beta I_S) C (\alpha + \beta I_S)$$

Then Schrodinger equation should correspond to equation governing changes in time of a $G_3^+$ - state. So, the actual difference between "classical" and "quantum" mechanics is only shift of viewpoint: instead of dealing with equations describing an axially symmetric physical object orientation, as in classical mechanics, we deal in quantum mechanics with equations describing evolution in time of the operation, $G_3^+$ - state, which returns the physical object orientation.

## 3.2. Stable states in terms of $G_3^+$ - states

In traditional quantum mechanics "state" of magnetic dipole in static magnetic field is defined by wave function $|\psi\rangle = z_1|0\rangle + z_2|1\rangle$ with the two basis "stable states" $|0\rangle$ and $|1\rangle$. In our terms the two "states" are, correspondingly, the set of all $G_3^+$ - states $\alpha + \beta_1 e_2 e_3$, with $\alpha^2 + \beta_1^2 = 1$, and the set of all $G_3^+$ - states $\beta_2 e_3 e_1 + \beta_3 e_1 e_2$, with $\beta_2^2 + \beta_3^2 = 1$. The bivector plane $e_2 e_3$ is taken as "complex plane"[6].

When any of the first kind of "stable $G_3^+$ - states" $\alpha + \beta_1 e_2 e_3$ acts on $e_2 e_3$ we get:

$$(\alpha - \beta_1 e_2 e_3) e_2 e_3 (\alpha + \beta_1 e_2 e_3) = (\alpha^2 + \beta_1^2) e_2 e_3 = e_2 e_3.$$

With the second kind, $\beta_2 e_3 e_1 + \beta_3 e_1 e_2$, we have:

$$(-\beta_2 e_3 e_1 - \beta_3 e_1 e_2) e_2 e_3 (\beta_2 e_3 e_1 + \beta_3 e_1 e_2) = -(\beta_2^2 + \beta_3^2) e_2 e_3 = -e_2 e_3.$$

This is actual meaning of quantum mechanical "stable states": any single one from the set of all $G_3^+$ - states, corresponding to $|0\rangle$, leaves $e_2 e_3$ unchanged, just rotates $e_2 e_3$ in its own plane. Every single $G_3^+$ - state from the set of all $G_3^+$ - states, corresponding to $|1\rangle$, flips $e_2 e_3$.

Let's take an arbitrary bivector $C = C_1 e_2 e_3 + C_2 e_3 e_1 + C_3 e_1 e_2$. Transforming it with $\alpha + \beta_1 e_2 e_3$ we get:

$$(\alpha - \beta_1 e_2 e_3) C (\alpha + \beta_1 e_2 e_3) = C_1 e_2 e_3 + [(\alpha - \beta_1^2) C_2 - 2\alpha\beta_1 C_3] e_3 e_1 + [(\alpha - \beta_1^2) C_3 + 2\alpha\beta_1 C_2] e_1 e_2$$

Parameterization $\alpha = \cos\varphi$, $\beta_1 = \sin\varphi$ gives:

$$(\alpha - \beta_1 e_2 e_3) C (\alpha + \beta_1 e_2 e_3) = C_1 e_2 e_3 + (C_2 \cos 2\varphi - C_3 \sin 2\varphi) e_3 e_1 + (C_2 \sin 2\varphi + C_3 \cos 2\varphi) e_1 e_2, \quad (3.3)$$

which means that the bivector component in $e_2 e_3$ does not change, while bivector $C_2 e_3 e_1 + C_3 e_1 e_2$ is rotated by angle $2\varphi$ around the axis of intersection of $e_3 e_1$ and $e_1 e_2$.

Taking a $G_3^+$ - state from the set of second kind basis $G_3^+$ - states $\beta_2 e_3 e_1 + \beta_3 e_1 e_2$ we get the result:

$$(-\beta_2 e_3 e_1 - \beta_3 e_1 e_2) C (\beta_2 e_3 e_1 + \beta_3 e_1 e_2) =$$
$$-C_1 e_2 e_3 + [(\beta_2^2 - \beta_3^2) C_2 + 2\beta_2\beta_3 C_3] e_3 e_1 + [(\beta_3^2 - \beta_2^2) C_3 + 2\beta_2\beta_3 C_2] e_1 e_2$$

Again, using parameterization $\beta_2 = \cos\vartheta$, $\beta_3 = \sin\vartheta$, we get:

---

[6] It does not actually matter which bivector in 3D is taken.

$$(-\beta_2 e_3 e_1 - \beta_3 e_1 e_2) C (\beta_2 e_3 e_1 + \beta_3 e_1 e_2) =$$
$$- C_1 e_2 e_3 + (C_2 \cos 2\vartheta + C_3 \sin 2\vartheta) e_3 e_1 + (C_2 \sin 2\vartheta - C_3 \cos 2\vartheta) e_1 e_2 \tag{3.4}$$

In addition to flipping the $e_2 e_3$ component of $C$, (3.4) shows that $\beta_2 e_3 e_1 + \beta_3 e_1 e_2$ rotates $C_2 e_3 e_1 + C_3 e_1 e_2$ by $2\vartheta$ around the axis of intersection of $e_3 e_1$ and $e_1 e_2$, but in opposite direction compared to the case of $G_3^+$ - state of the first kind, $\alpha + \beta_1 e_2 e_3$.

Let's take $G_3^+$ - state written in arbitrary bivector basis $\{B_1, B_2, B_3\}$, $B_1 B_2 = -B_3$, $B_1 B_3 = B_2$, $B_2 B_3 = -B_1$. The two above states then correspond to two sets of $G_3^+$ - states:

$$|0\rangle \leftrightarrow \{\forall (\alpha + \beta_1 B_1): \alpha^2 + \beta_1^2 = 1\}$$

$$|1\rangle \leftrightarrow \{\forall (\beta_2 B_2 + \beta_3 B_3): \beta_2^2 + \beta_3^2 = 1\}$$

If an arbitrary bivector is expanded in basis $\{B_1, B_2, B_3\}$, $C = C_1 B_1 + C_2 B_2 + C_3 B_3$, then for $G_3^+$ - states from the classes of equivalence $|0\rangle$ or $|1\rangle$ we have, using (2.3) with $e_2 e_3 \to B_1$, $e_3 e_1 \to B_2$, $e_1 e_2 \to B_3$,

$$(\alpha - \beta_1 B_1) C (\alpha + \beta_1 B_1) = C_1 B_1 + [C_2 (\alpha^2 - \beta_1^2) - 2 C_3 \alpha \beta_1] B_2 + [C_3 (\alpha^2 - \beta_1^2) + 2 C_2 \alpha \beta_1] B_3 =$$
$$C_1 B_1 + (C_2 \cos 2\varphi - C_3 \sin 2\varphi) B_2 + (C_2 \sin 2\varphi + C_3 \cos 2\varphi) B_3$$

$$(-\beta_2 B_2 - \beta_3 B_3) C (\beta_2 B_2 + \beta_3 B_3) = -C_1 B_1 + [C_2 (\beta_2^2 - \beta_3^2) + 2 C_3 \beta_2 \beta_3] B_2 + [2 C_2 \beta_2 \beta_3 - C_3 (\beta_2^2 - \beta_3^2)] B_3 =$$
$$- C_1 B_1 + (C_2 \cos 2\vartheta + C_3 \sin 2\vartheta) B_2 + (C_2 \sin 2\vartheta - C_3 \cos 2\vartheta) B_3$$

The bottom line is:

- When using any $G_3^+$ - state $\alpha + \beta_1 B_1, \alpha^2 + \beta_1^2 = 1$ from the equivalence class $|0\rangle$ to observe the result of transformed bivector $C = C_1 B_1 + C_2 B_2 + C_3 B_3$, the component $C_1$ remains unchanged, though $C$ rotates around the axis orthogonal to $B_1$ by angle $2\varphi$ defined by $\alpha = \cos \varphi$ and $\beta_1 = \sin \varphi$; the $C$ projection components onto $B_2$ and $B_3$ get changed as rotated by the same angle around the same axis.

- When using any $G_3^+$ - state $\beta_2 e_3 e_1 + \beta_3 e_1 e_2$, $\beta_2^2 + \beta_3^2 = 1$ from the equivalence class $|1\rangle$ to observe the result of transformed bivector $C = C_1 B_1 + C_2 B_2 + C_3 B_3$, the component $C_1$ flips, $C_1 \to -C_1$, though no rotation of $C$ around axis orthogonal to $B_1$ is detected; the $C$ projection components onto $B_2$ and $B_3$ get transformed as rotated by angle $2\vartheta$ defined by $\beta_2 = \cos \vartheta$, $\beta_3 = \sin \vartheta$ around the axis orthogonal to $B_1$ but rotation direction is opposite to the first case of $|0\rangle$.

In the case of $|0\rangle$ rotation in $B_1$ plane "drags" rotations of two other bivector $C$ projections. In the case of $|1\rangle$ $B_2$ and $B_3$ components rotate, while any vector in the $B_1$ component remains fixed.

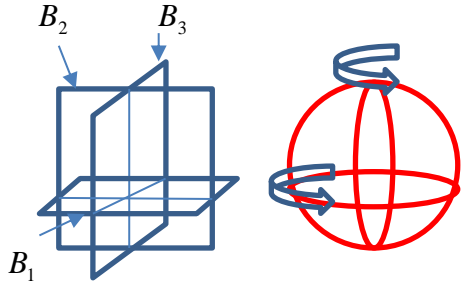

The case of rotation of bivector with $G_3^+$ - state from $|0\rangle$ in $B_1$ plane that causes rotations of bivector $B_2$ and $B_3$ components by the same angle.

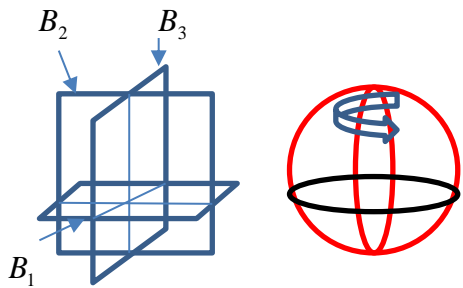

The case of rotation of bivector with $G_3^+$ - state from $|1\rangle$ causes rotation of bivector $B_2$ and $B_3$ components by corresponding angle, though all vectors in $B_1$ remain immovable.

The "dragging" phenomenon will be considered in the next subsection in the connection with Clifford translations and quantum mechanical state equivalence transformations $|\psi\rangle \to e^{i\theta}|\psi\rangle$.

## 3.3. Clifford translations in terms of $G_3^+$ - states

In quantum mechanics two state system wave functions are considered equivalent when related by $|\psi\rangle \to e^{i\theta}|\psi\rangle$[7]. This is not true, as was shown in previous subsection, for $G_3^+$ - states which give more detailed information because are associated with explicitly defined "complex" planes in 3D.

Let's look at Clifford translations in terms of $G_3^+$ - states of the class

$$|0\rangle \leftrightarrow \{\forall(\alpha + \beta_1 B_1): \alpha^2 + \beta_1^2 = 1\}$$

in dynamical situation when phase $\theta$ is changing in time.

---

[7] This type of transformation is often called Clifford translations.

Clifford translations should accurately be written in 3D as $so(\alpha, \vec{\beta}) \to e^{I_s \theta} so(\alpha, \vec{\beta})$. Again, I will initially consider, to simplify calculations, the cases when unit bivector $I_S$ lies in one of the coordinate planes: $I_S = e_2 e_3$, $I_S = e_3 e_1$ or $I_S = e_1 e_2$.

If $I_S = e_2 e_3$ then "velocity" of Clifford translation is:

$$\frac{d}{dt}\left[e^{e_2 e_3 \theta} so(\alpha, \vec{\beta})\right] = \frac{d}{dt}[\cos\theta + e_2 e_3 \sin\theta] so(\alpha, \vec{\beta}) = [(\sin\theta) e_2 e_3 e_2 e_3 + (\cos\theta) e_2 e_3] so(\alpha, \vec{\beta}) \theta' =$$
$$e^{e_2 e_3 \theta} e_2 e_3 so(\alpha, \vec{\beta}) \theta'$$

This $G_3^+$ - state is orthogonal to the Clifford translation orbit $e^{e_2 e_3 \theta} so(\alpha, \vec{\beta})$ and spans one-dimensional vertical subspace of the Clifford orbit tangent space.

Two more $G_3^+$ - states, $e^{e_2 e_3 \theta} e_3 e_1 so(\alpha, \vec{\beta}) \theta'$, $e^{e_2 e_3 \theta} e_1 e_2 so(\alpha, \vec{\beta}) \theta'$, are orthogonal to $e^{e_2 e_3 \theta} so(\alpha, \vec{\beta})$, $e^{e_2 e_3 \theta} e_2 e_3 so(\alpha, \vec{\beta}) \theta'$ and to each other. They span two-dimensional horizontal subspace of the Clifford orbit tangent space.

Using other two planes for Clifford translations, $I_S = e_3 e_1$ and $I_S = e_1 e_2$, we get correspondingly vertical tangents $e^{e_3 e_1 \theta} e_3 e_1 so(\alpha, \vec{\beta}) \theta'$ and $e^{e_1 e_2 \theta} e_1 e_2 so(\alpha, \vec{\beta}) \theta'$, and couples of horizontal tangents

$\{e^{e_3 e_1 \theta} e_2 e_3 so(\alpha, \vec{\beta}) \theta'$, $e^{e_3 e_1 \theta} e_1 e_2 so(\alpha, \vec{\beta}) \theta' \}$, $\{e^{e_1 e_2 \theta} e_3 e_1 so(\alpha, \vec{\beta}) \theta'$, $e^{e_1 e_2 \theta} e_2 e_3 so(\alpha, \vec{\beta}) \theta' \}$.

Generalizing the above results to an arbitrary great circle plane spanned by unit bivector $I_S$ of Clifford translation $so(\alpha, \vec{\beta}) \to e^{I_s \theta} so(\alpha, \vec{\beta})$, we have the following:

- the vertical tangent of the translation (speed along fiber) is $e^{I_s \theta} I_S so(\alpha, \vec{\beta}) \theta'$
- two tangents, spanning the horizontal subspace of the tangent space, are
  $e^{I_s \theta} I_{S_1} so(\alpha, \vec{\beta}) \theta'$ and $e^{I_s \theta} I_{S_2} so(\alpha, \vec{\beta}) \theta'$, where $S_1$ and $S_2$ are two planes in 3D orthogonal to $S$ and to each other.

1. Conclusions
    - The two state wave functions $|\psi\rangle = (c_1, c_2)^T$, where the components $c_1, c_2$ of the column are "complex" numbers, are objects with lost information about actual physical states.
    - Loosing of information happened due to formal using of "imaginary units" in pure algebraic way.

- "States" should really be elements $so(\alpha, \beta, S) \equiv \alpha + \beta I_S =$
  $\alpha + \beta(b_1 e_2 e_3 + b_2 e_3 e_1 + b_3 e_1 e_2)$, $\alpha^2 + \beta^2 = 1$, $b_1^2 + b_2^2 + b_3^2 = 1$, of even geometrical algebra $G_3^+$, unambiguously representing rotations in 3D.
- Actual difference between "classical" and "quantum" mechanics is only shift of viewpoint: instead of dealing with equations describing an axially symmetric physical object orientation, as in classical mechanics, quantum mechanics deals with equations describing evolution in time of the operation, $G_3^+$ - state, which returns the physical object orientation.

## Works Cited

[1] D. Hestenes, New Foundations of Classical Mechanics, Dordrecht/Boston/London: Kluwer Academic Publishers, 1999.

[2] C. Doran and A. Lasenby, Geometric Algebra for Physicists, Cambridge: Cambridge University Press, 2010.

[3] A. Soiguine, "A tossed coin as quantum mechanical object," September 2013. [Online]. Available: http://arxiv.org/abs/1309.5002.

[4] A. Soiguine, Vector Algebra in Applied Problems, Leningrad: Naval Academy, 1990 (in Russian).

[5] A. Soiguine, "Complex Conjugation - Realtive to What?," in *Clifford Algebras with Numeric and Symbolic Computations*, Boston, Birkhauser, 1996, pp. 285-294.

[6] J. v. Neumann, Mathematical Foundations of Quantum Mechanics, Princeton: Priceton University Press, 1955.